\documentclass[conference,10pt]{IEEEtran}
\usepackage[utf8]{inputenc}
\IEEEoverridecommandlockouts

\usepackage{verbatim}
\usepackage{color}
\usepackage{multicol}
\usepackage{placeins}
\usepackage{float}
\usepackage{graphicx}
\usepackage{cite}

\usepackage[caption=false,font=footnotesize]{subfig}

\usepackage{xifthen}
\usepackage{xcolor} 

\newcommand\alternative[2][1]{
	\ifthenelse{\equal{#1}{1}}{%
		\textcolor{blue}%
	}{%
		\ifthenelse{\equal{#1}{2}}{
			\textcolor{green}%
		}{%
			\textcolor{magenta}%
	}}
	{\textbf{#2}} 
	\PackageWarning{ALTERNATIVE:}{ALT: #2}
}

\usepackage{pgf}	
\usepackage{pgfplots}
\usepackage{tikz}
\usetikzlibrary{decorations.pathreplacing}
\usetikzlibrary{plotmarks}
\usepackage{pgfplotstable}

\usetikzlibrary{external}
\usetikzlibrary{spy}
\pgfplotsset{compat=1.4}

\usepackage{algorithmic}
\usepackage{algorithm}

\usepackage{array}
\newcolumntype{L}[1]{>{\raggedright\let\newline\\\arraybackslash\hspace{0pt}}m{#1}}
\newcolumntype{C}[1]{>{\centering\let\newline\\\arraybackslash\hspace{0pt}}m{#1}}
\usepackage{booktabs}

\usepackage{amsmath}
\usepackage{amssymb}
\usepackage{amsthm}
\usepackage{bm}

\DeclareMathOperator{\EX}{\mathbb{E}} 
\newcommand{\transpose}{^{\mkern-1.5mu\mathsf{T}}}
\newcommand{\hermitian}{^{\mathsf{H}}} 
\newcommand{\conj}{^*} 

\DeclareMathOperator{\diag}{diag}

\DeclareMathOperator{\I}{\mathsf{j}}

\newcommand{\norm}[1]{\left\lVert #1 \right\rVert}

\newcommand{\reals}{\mathbb{R}}
\newcommand{\complexs}{\mathbb{C}}

\newcommand{\CN}[2]{\mathcal{CN}\left( #1,#2 \right)}

\newcommand{\grad}{\bm{\theta}}
\newcommand{\diffgrad}{\Delta\bm{\theta}}

\newcommand{\A}{\bm{A}}
\newcommand{\x}{\bm{x}}

\newcommand{\Y}{\bm{Y}}

\newcommand{\Noise}{\bm{N}}

\newcommand{\tauc}{\tau_{\text{c}}}
\newcommand{\taup}{\tau_{\text{p}}}
\newcommand{\channel}{\bm{g}}
\newcommand{\nchannel}{\bm{h}}
\newcommand{\Channels}{\bm{G}}
\newcommand{\pilot}{\bm{\phi}}
\newcommand{\betamin}{\beta_\text{min}}

\title{Optimal MIMO Combining for Blind Federated Edge Learning with Gradient Sparsification}
\author{\IEEEauthorblockN{Ema Becirovic,  Zheng Chen, and Erik G. Larsson}
	\IEEEauthorblockA{Dept. of Electrical Engineering (ISY), Link\"oping University, Link\"oping, Sweden \\
		Email: \{ema.becirovic, zheng.chen, erik.g.larsson\}\makeatletter @\makeatother liu.se
	}\thanks{This work was supported in part by KAW foundation and in part by ELLIIT.}
	\thanks{The computations were enabled by resources provided by the National
		Supercomputer Centre (NSC), funded by Linköping University.}
}

\begin{document}
	\maketitle
	
	\begin{abstract}
		We provide the optimal receive combining strategy for federated learning in multiple-input multiple-output (MIMO) systems. Our proposed algorithm allows the clients to perform individual gradient sparsification which greatly improves performance in scenarios with heterogeneous (non i.i.d.) training data. The proposed method beats the benchmark by a wide margin.
		
	\end{abstract}
	
	
	\section{Introduction} 
	Federated learning (FL) is a technology where a set of distributed clients, possessing individual training data, can keep their data privacy while cooperatively training a machine learning model with the assistance of a parameter server \cite{McMahan17}. 
	In essence, FL can be viewed as one instance of distributed stochastic gradient descent and works in a two-step process. In the first step, the clients train a machine learning model based on their local data and transmit the learned parameters to the parameter server.  In the second step, the parameter server aggregates the parameters from each client and broadcasts the aggregated model back to the clients. This process continues until the model converges.

	The acquisition of the clients' parameters can be done in many different ways.
	Recently, over-the-air computation (OtA) has gained the spotlight as a promising data aggregation scheme that uses the superposition property \cite{Goldenbaum15} of the wireless channels \cite{Wei22,Adeli22,Sahin21,Amiri21bfeel,Chen22}.
	It has been shown that 
	without much performance degradation, the gradient updates can be heavily quantized and sparsified \cite{Seide14,Alistarh17,Bernstein18,Strom15,Aji17,Li21}, when the goal is to compute the weighted average of the gradient updates rather than the exact gradient values from each client.
	
	In an OtA FL system, if the clients have perfect channel knowledge, before transmitting the model updates, they can pre-process the data 
	to compensate for the channel gain and phase, to constructively add up the received signals at the base station/parameter server. 
	In the case of no channel state information at the clients, one option is to use non-coherent transmission techniques to infer the aggregated gradient \cite{Sahin21,Adeli22}.
	Another option is that the clients transmit pilots from which the parameter server can estimate the channels and design a receive combining method in similar fashion as in massive MIMO \cite{redbook}. 	
	What has been proposed in the literature \cite{Amiri21bfeel,Wei22} is for the clients to send the \emph{same pilot} which effectively means that the parameter server can only estimate the \emph{sum channel} of the clients. 	
	The advantage of estimating the sum channel from a single pilot is that the channel estimation error, of the sum channel, will be smaller than it estimating the clients' channels separately and then adding them up. The pilot overhead can be made smaller than when estimating the individual channels and hence, estimating the sum channel has been advocated.

	\textbf{Contributions:} We provide the optimal receive combining strategy for MIMO systems without channel knowledge at the clients. This is done by using the best linear unbiased estimator (BLUE) of the transmitted signals after obtaining individual channel estimates. We show that having individual channel estimates gives significant gains. Firstly, the clients can independently perform power control. Secondly, and more importantly, the clients can independently quantize and encode their data (e.g., with sparsification), which is impossible with standard OtA \cite{Amiri21bfeel,Wei22}. 
	Simulation results show that the proposed algorithm beats the benchmark in \cite{Wei22,Amiri21bfeel} by a wide margin.

	\section{System Model}
	We consider an FL system with $K$ single-antenna clients participating in training a global learning model assisted by an edge server (base station) equipped with $M$ antennas. The learning model is represented by a $d$-dimensional parameter vector $\grad\in\reals^d$. The goal is to find the optimal parameter vector $\grad^*$ that minimizes an empirical loss function defined by
	\begin{equation}
	F(\grad)=\textstyle\sum_{k=1}^{K}w_k F_k(\grad),
	\label{eq:globalLoss_f}
	\end{equation}    
	where $F_k(\grad)$ is the loss function computed over the local training data set possessed by client $k$, and $w_k$ is the weight factor associated with client $k$.
	
	\subsection{Federated Averaging}
	The most representative FL framework is Federated Averaging \cite{McMahan17}. In the $t$:th communication round with $t=1,2,\ldots$, the following steps are executed:
	\begin{enumerate}
		\item The server broadcasts the current global model $\grad(t)$.
		\item Each client $k$ runs stochastic gradient descent (SGD) and the update rule follows
		\begin{equation}
		\grad_{k}(t+1)=\grad_{k}(t)-\alpha_t\nabla F_k(\grad_{k}(t)),
		\label{eq:localItUpdate}
		\end{equation}
		where $\alpha_t$ is the step size at iteration $t$.
		Let $\diffgrad_k(t)$ represent the local model update from client $k$: $\diffgrad_k(t)=\grad_{k}(t+1)-\grad_{k}(t)$.\footnote{The clients can run several steps of local SGD within each communication round. For simplicity, we omit the local iteration index here.}
		\item After receiving the local updates from the clients, the server aggregates the received information, 
		\begin{equation}
		\diffgrad(t) = \textstyle\sum_{k=1}^Kw_k\diffgrad_k(t),
		\end{equation} and updates the global model as
		\begin{equation}
		\grad(t+1)=\grad(t)+\alpha_t\diffgrad(t).
		\label{eq:syncFlAggregation}	
		\end{equation}
	\end{enumerate}
	
	In this work, we focus on the transmission and aggregation of the gradient updates from the client to the parameter server.

	\subsection{Channel Assumptions}
	For the wireless channels between the base station and the clients, we assume a block fading model (applicable for both wide- and narrowband systems), where in each coherence interval the channel is time invariant and frequency flat. The channel between client $k$ and the base station is modeled as i.i.d. Rayleigh fading, $\channel_k \sim \CN{\bm{0}}{\beta_k\bm{I}}, $
	where $\beta_k$ is the large-scale fading coefficient. The normalized channel, which only captures small-scale fading effects, is denoted by $\nchannel_k = \frac{1}{\sqrt{\beta_k}}\channel_k$.
	We let $\Channels = \left[ \channel_1,\dots,\channel_K\right]$.
	The channels of different clients are uncorrelated, $\EX\left\{ \channel_k\channel_{k'}\hermitian \right\} = \bm{0}$, $k\neq k'$. We assume that the coherence interval is $\tauc$ samples long. 
	
	\section{Channel Estimation}
	In each coherence interval, the base station estimates the channels from pilots transmitted by the clients. The pilots are all $\taup < \tauc$ symbols long. 
	We consider two types of pilot transmission.
	
	\subsection{Orthogonal pilots}
	With the first type of pilot transmission, we aim to estimate all the users' channels. The users transmit mutually orthogonal pilots. 
	The base station receives
	\begin{align}
	\Y_\text{p,orth} &= \textstyle\sum_{k=1}^{K} \sqrt{\rho \taup} \channel_k \pilot_k\hermitian + \Noise_\text{p},
	\end{align}
	where $\rho$ is the signal-to-noise ratio (SNR), $\sqrt{\taup}\pilot_k\conj$ is the pilot for client $k$ and $\norm{\pilot_k}_2^2=1$ and $\pilot_k\hermitian\pilot_{k'} = 0$, $k\neq k'$, and $\Noise_\text{p} \in \complexs^{M\times\taup}$ is  noise with independent $\CN{0}{1}$ elements.
	
	To estimate the channel for client $k$, we use the sufficient statistic $\Y_\text{p,orth}\pilot_k = \sqrt{\rho \taup}\channel_k + \Noise_\text{p}\pilot_k $ to obtain the minimum mean-square error (MMSE) channel estimate \cite[Ch.~3]{redbook}\cite[Ch.~10]{Kay93estimation}
	\begin{equation}
	\hat{\channel}_k = \frac{\sqrt{\rho \taup }\beta_k}{1+\rho\taup\beta_k}\Y_\text{p,orth}\pilot_k. \label{eq:orth-channel-estimate}
	\end{equation}
	The mean-square of the channel estimate per antenna is 
	\begin{equation}
	\gamma_k  = \frac{\rho\taup\beta_k^2}{1+\rho\taup\beta_k}.
	\end{equation}
	The channel estimation error is $\tilde{\channel}_k = \hat{\channel}_k - \channel_k $
	and it is uncorrelated with both $\channel_k$ and $\hat{\channel}_k$. 
	We let $\hat{\Channels} = \left[\hat{\channel}_1,\dots,\hat{\channel}_K\right]$.
	
	\subsection{Same pilot}
	With this pilot transmission, we aim to estimate the sum of normalized client channels, i.e., $\nchannel_{\text{sum}} = \sum_{k=1}^K\nchannel_k$. All the clients transmit the same pilot, $\sqrt{\taup}\pilot\conj$ and they scale their transmitted power with $\betamin/\beta_k$, $k=1,\dots,K$, where $\betamin=\min_k \beta_k$, such that they effectively transmit over the normalized channels. The base station receives
	\begin{equation}
	\resizebox{1\hsize}{!}{$
		\Y_\text{p,sum} = \displaystyle\sum_{k=1}^{K} \sqrt{\rho \taup \frac{\betamin}{\beta_k}} \channel_k \pilot\hermitian + \Noise_\text{p} = \displaystyle\sum_{k=1}^{K} \sqrt{\rho \taup \betamin}  \nchannel_k \pilot\hermitian + \Noise_\text{p}.$}
	\end{equation}
	This gives $\Y_\text{p,sum}\pilot =  \sqrt{\rho \taup \betamin} \textstyle\sum_{k=1}^{K}\nchannel_k  + \Noise_\text{p}\pilot$.
	We get the MMSE estimate
	\begin{align}
	\hat{\nchannel}_\text{sum} = \frac{\sqrt{\rho \taup \betamin}K}{1 + \rho\taup\betamin K}\Y_\text{p,sum}\pilot.\label{eq:sum-channel-estimate}
	\end{align}
	The mean-square of the channel estimate is 
	\begin{equation}
	\bar{\gamma} = \frac{\rho\taup\betamin K^2}{1+\rho\taup\betamin K}.
	\end{equation}

	Note that, this method estimates the sum channel better than adding up the individual channel estimates in \eqref{eq:orth-channel-estimate}. To see this, consider a case where the clients have equal channel quality, say $\beta_k = \beta = \betamin$. The mean-square error (MSE) per antenna of the estimated sum channel from orthogonal pilots, obtained by summing \eqref{eq:orth-channel-estimate}, is
	\begin{equation}
	K\beta -{\textstyle\sum_{k=1}^{K}\gamma_k}= \frac{K\beta}{1+\rho\taup\beta}.
	\label{eq:mse-orth}
	\end{equation}
	The MSE of the estimated sum channel from the same pilot, obtained from \eqref{eq:sum-channel-estimate}, is
	\begin{equation}
	K\beta - \bar{\gamma} = \frac{K\beta}{1+\rho\taup\beta K},
	\end{equation}
	which is approximately a factor $K$ smaller than \eqref{eq:mse-orth}.
	
	\section{Data Transmission and Combining}
	\label{sec:transmission}
	After each round of local training, the goal is to compute $\diffgrad(t)$ at the edge server.
	Before transmitting the model updates,
	at each client $k$, the (real) model update is split in half to make a (complex) vector,
	\begin{equation}
	\x^\text{full}_k = \operatorname{SPLIT}(\diffgrad_k(t)), \label{eq:split}
	\end{equation}
	i.e., the $i$:th component is\footnote{For simplicity, we assume that $d$ is a even number.}
	\begin{equation}
	\left[\x_k^\text{full}\right]_i =  (\left[\diffgrad_k\right]_i + \I \left[\diffgrad_k\right]_{i+d/2}),  \quad i = 1,\dots,d/2. \label{eq:transmit-vector}
	\end{equation}
	Due to limited communication resources, $\x^{\text{full}}_k$ is sparsified and multiplied by a $T\times d/2$ measurement matrix, $\A_k^t$, where $T \ll d/2$ is the number of transmitted analog samples. $\A_k^t$ can be any matrix but some properties allow for efficient estimation of the sparse vector, for example the restricted isometry property (RIP) \cite{Blumensath09}. With a certain probability, RIP can be achieved with e.g., random matrices with i.i.d. Gaussian entries \cite{Rish14}.
	
	The vector transmitted by client $k$ is
	\begin{align}
	\sqrt{\eta_k}  \x_k = \sqrt{\eta_k}\A_k^t \x_k^\text{sparse},
	\end{align}
	where $\x_k = \A_k^t\x_k^\text{sparse}$, 
	and 
	$\eta_k \geq 0$ is the power control coefficient of client $k$. The power control coefficient is chosen such that $\norm{\sqrt{\eta_k}\x_k}_2^2\leq T$. Additionally, 
	\begin{equation}
	\x_k^\text{sparse} = \operatorname{SPARSE}(\x_k^\text{full} + \bm{r}_k,S),
	\end{equation}
	where $S$ is the sparsity level, and $\bm{r}_k$ is the residual from the  sparsification in the \emph{previous} global iteration. Using superscripts denoting time,  $\bm{r}_k^{t+1} = \x^\text{full}_k + \bm{r}_k^{t} - \x^\text{sparse}_k$, i.e., the residual accumulates over time \cite{Aji17}.
	If $\taup + T\leq \tauc$ we can send the whole model update in the same coherence interval which is what we assume in our experiments in Section~\ref{sec:numerical-results}. However, nothing in principle prevents the splitting of the model over multiple coherence intervals.
	
	The base station (edge server) receives
	\begin{equation}
	\Y =\textstyle \sum_{k=1}^{K}  \sqrt{\rho\eta_k}\channel_k \bm{x}_k\transpose + \Noise,  \label{eq:data-transmission}
	\end{equation}
	where  $\rho$ is the signal-to-noise ratio (SNR) and $\Noise$ is noise with independent $\CN{0}{1}$ entries.

	The design of the receive combining vector depends on whether we have estimates of the individual channels or the sum channel. 
	\subsection{Proposed Sparse BLUE}
	If the base station has individual channel estimates, we propose using the BLUE\footnote{Also known as zero-forcing combining.}. The estimates of the transmitted signals are
	\begin{equation}
	[\hat{\x}_1,\dots,\hat{\x}_k] = \left(\frac{1}{\sqrt{\rho}}\bm{D}_{\eta}^{-1/2}(\hat{\Channels}\hermitian\hat{\Channels})^{-1}\hat{\Channels}\hermitian\Y\right)\transpose,
	\end{equation}
	\mbox{where $\bm{D}_\eta = \diag(\eta_1,\dots,\eta_K)$. 
		The estimates are unbiased, i.e.,}
	\begin{equation}
	\EX\left\{\hat{\x}_k \mid \hat{\Channels} \right\} = \x_k = \A_k^t\x_k^\text{sparse}.
	\end{equation}
	An estimate of the sparsified gradient of client $k$ can then be found by solving 
	\begin{align}
	\hat{\x}^\text{sparse}_k = &\min_{\x} \norm{\A_k^t\x - \hat{\x}_k}_2^2 \label{eq:opt}\\
	&\text{s.t.} \norm{\x}_0 \leq S. \nonumber 
	\end{align} 
	After solving the optimization problems, 
	the splitting process in \eqref{eq:split} is undone, 
	\begin{equation}
	\widehat{\diffgrad}_k(t) = \operatorname{UNSPLIT}(\hat{\x}_k^\text{sparse}),
	\end{equation}
	and the estimated gradients are aggregated, $\widehat{\diffgrad}(t) = \sum_{k=1}^K w_k \widehat{\diffgrad}_k$.  Finally, the global model is updated, 
	\begin{equation}
	\grad(t+1) = \grad(t) + \alpha_t\widehat{\diffgrad}(t). \label{eq:update-est}
	\end{equation}
	\mbox{The complete proposed algorithm is summarized in Algorithm~\ref{alg:sparse-blue}.}

	\subsection{Benchmark with Sum Channel Estimate \cite{Amiri21bfeel,Wei22}}
	If the base station only has the sum channel estimate we multiply the received signal by the conjugate of the estimate similarly to what is done in \cite{Amiri21bfeel,Wei22}, to obtain an estimate of $\x = \sum_{k=1}^Kw_k\A_k^t\x_k^\text{sparse}$,
	\begin{equation}
	\hat{\x} = c \,(\hat{\nchannel}_\text{sum}\hermitian\Y)\transpose, \label{eq:sum-estimate}
	\end{equation} 
	where $c$ is a scaling constant.
	However, this scheme has some restrictions: 
	\begin{enumerate}
		\item The received power at the base station needs to be the same for all clients and the weights of the gradients need to be scaled by the clients, i.e., $\eta_k = \eta\frac{w_k^2}{\beta_k}$ where $\eta$ is chosen such that 
		\begin{equation}
		\max_k \norm{\sqrt{\eta_k}\x_k}_2^2 = \max_k \eta\frac{w_k^2}{\beta_k} \norm{\x_k}_2^2 = T.
		\label{eq:power-constraint-sum}
		\end{equation}
		This requires that the base station has knowledge of $\norm{\x_k}_2^2$ for all clients, which either needs to be signaled from each client or predicted at the base station.
		In contrast, in the proposed sparse BLUE scheme all clients can individually control their powers. Especially, they can transmit with full power, $\eta_k = \frac{T}{\norm{\x_k}_2^2}$.
		The power control is ``slow'' in the sense that it is not a function of the small-scale fading, but it needs to adapt to the gradients for each transmission.
		\item All clients need to use the same measurement matrix in each iteration, i.e., $\A_k^t =  \A^t$. 
	\end{enumerate}
	We choose\footnote{The scaling factor differs from \cite{Amiri21bfeel,Wei22} but this does not impact the analysis since the scaling factor can be absorbed into the step size, $\alpha_t$, which, as shown later, is selected to ensure a fair comparison.} $c = \frac{K}{M\sqrt{\eta\rho}\bar{\gamma}}$,
	since on average 
	\begin{align}
	\EX\left\{ \hat{\x} \right\} &= \EX\left\{ \frac{K (\hat{\nchannel}_{\text{sum}}\hermitian\Y)\transpose}{M\sqrt{\eta\rho}\bar{\gamma}} \right\} = \A^t\sum_{k=1}^K w_k\x_k^\text{sparse} ,
	\end{align}
	where the expectation is over channel realizations and noise. Hence, we can use the same step size to fairly compare the sparse BLUE method to this method. However, note that, the estimate \eqref{eq:sum-estimate} is \emph{not} unbiased, since in general $
	\EX\left\{ \hat{\x} \mid \hat{\nchannel}_\text{sum} \right\} \neq \A^t\sum_{k=1}^{K}w_k\x_k^\text{sparse}$.
	
	We proceed by solving 
	\begin{align}
	\widehat{\left(\sum_{k=1}^Kw_k\x_k^\text{sparse}\right)} = &\min_{\x} \norm{\A^t\x - \hat{\x}}_2^2 \label{eq:optsum}\\
	&\text{s.t.} \norm{\x}_0 \leq G, \nonumber 
	\end{align} 
	where $G=S$ if the clients used the same sparsity pattern (which needs to be coordinated somehow) and $G=KS$ if the clients used different sparsity patterns. If the clients coordinate the sparsity pattern, at most $S$ non-zero components need to be estimated, while if the clients choose different sparsity patterns, there can be $KS$ non-zero components.
	After solving the problem the splitting process in \eqref{eq:split} is undone,
	\begin{equation}
	\widehat{\diffgrad}(t) = \operatorname{UNSPLIT}\left(\widehat{\left(\sum_{k=1}^Kw_k\x_k^\text{sparse}\right)}\right)
	\end{equation} and the global model is updated, as in \eqref{eq:update-est}.
	
	The complete algorithm is summarized in Algorithm~\ref{alg:sparse-sum}. Note that Algorithm~\ref{alg:sparse-sum} has a lower complexity than Algorithm~\ref{alg:sparse-blue} since only one sparsity problem is solved.
	
	\begin{figure}[tbp]
		\begin{algorithm}[H]
			\algsetup{indent=1em}
			\begin{algorithmic}[1]
				\FOR{global iteration $t\in\{1,2,\dots\}$}			
				\FOR{client $k \in \{1,\dots,K\}$ \textbf{in parallel}} 
				\STATE receive (error free) $\grad(t)$ \label{alg:blue:line:a}
				\STATE obtain $\diffgrad_k(t)$ from stochastic gradient descent
				\STATE $\x_k^\text{full} \gets \operatorname{SPLIT}(\diffgrad_k(t))$
				\STATE $\x_k^\text{full} \gets \x_k^\text{full} + \bm{r}_k$
				\STATE $\x^\text{sparse}_k \gets \operatorname{SPARSE}(\x_k^\text{full},S)$
				\STATE $\bm{r}_k \gets \x_k^\text{full} - \x_k^\text{sparse}$, initially $\bm{0}$ \label{alg:blue:line:b}
				\STATE $\x_k \gets \bm{A}_k^t \x_k^\text{sparse}$ 
				\STATE transmit $\pilot_k$, $\sqrt{\eta_k}\x_k$	
				\ENDFOR
				
				\textbf{The parameter server/base station does:}
				\STATE receive $\Y_\text{p,orth}$ and $\Y$
				\STATE estimate $\hat{\Channels}$ through \eqref{eq:orth-channel-estimate}
				\FOR{$k\in\{1,\dots,K\}$}
				\STATE $\hat{\x}_k \gets \frac{1}{\sqrt{\eta_k\rho}}[\hat{\Channels}(\hat{\Channels}\hermitian\hat{\Channels})^{-1}]_k\hermitian\Y$
				\STATE solve \eqref{eq:opt} to get $\hat{\x}_k^\text{sparse}$
				\STATE $\widehat{\diffgrad}_k(t,\tau) \gets \operatorname{UNSPLIT}(\hat{\x}_k^\text{sparse})$
				\ENDFOR
				\STATE $\widehat{\diffgrad}(t) \gets  \sum_{k=1}^Kw_k\widehat{\diffgrad}_k(t)$ 
				\STATE $\grad(t+1) \gets \grad(t) + \alpha_t^\text{global}\widehat{\diffgrad}(t)$\label{alg:blue:line:c}
				\STATE broadcast (error free) $\grad(t+1)$ \label{alg:blue:line:d}
				\ENDFOR
			\end{algorithmic}
			\caption{Sparse BLUE algorithm}
			\label{alg:sparse-blue}
		\end{algorithm}
		\vspace{-3em}
	\end{figure}
	
	\begin{figure}[tbp]
		\begin{algorithm}[H]
			\algsetup{indent=1em}
			\begin{algorithmic}[1]
				\FOR{global iteration $t\in\{1,2,\dots\}$}			
				\FOR{client $k \in \{1,\dots,K\}$ \textbf{in parallel}} 
				\STATE same procedure as in Algorithm~\ref{alg:sparse-blue}, lines \ref{alg:blue:line:a}--\ref{alg:blue:line:b}.
				\STATE $\x_k \gets \bm{A}^t \x_k^\text{sparse}$ 
				\STATE transmit $\pilot$, $\sqrt{\eta_k}\x_k$	
				\ENDFOR
				
				\textbf{The parameter server/base station does:}
				\STATE receive $\Y_\text{p,sum}$ and $\Y$
				\STATE estimate $\hat{\nchannel}_\text{sum}$ through \eqref{eq:sum-channel-estimate}
				
				\STATE $\hat{\x} \gets  \frac{K}{M\sqrt{\rho\eta}\bar{\gamma}} \hat{\nchannel}_\text{sum}\hermitian\Y$
				\STATE solve \eqref{eq:optsum} to get $\widehat{\left(\sum_{k=1}^Kw_k\x_k^\text{sparse}\right)}$
				\STATE $\widehat{\diffgrad}(t) \gets \operatorname{UNSPLIT}(\widehat{\left(\sum_{k=1}^Kw_k\x_k^\text{sparse}\right)})$
				\STATE same procedure as in Algorithm~\ref{alg:sparse-blue}, lines \ref{alg:blue:line:c}--\ref{alg:blue:line:d}.
				\ENDFOR
			\end{algorithmic}
			\caption{Sparse SUM algorithm}
			\label{alg:sparse-sum}
		\end{algorithm}
		\vspace{-2.5em}
	\end{figure}

	\begin{table}[t]
		\centering 
		\caption{Network architecture used in the numerical results.}
		\label{tab:net}
		\renewcommand{\arraystretch}{1.3}
		\begin{tabular}{|c|}
			\hline 
			$3\times3$ convolutional layer, 16 filters, ReLU activation, no padding  \\ \hline 
			$2\times 2$ max pooling \\ \hline 
			$3\times3$ convolutional layer, 32 filters, ReLU activation, no padding  \\ \hline
			$2\times 2$ max pooling \\ \hline 
			Flatten \\ \hline
			Dense layer with $10$ outputs, softmax activation \\ \hline 
		\end{tabular}
		\renewcommand{\arraystretch}{1}
		\vspace{-1.5em}
	\end{table}

	\section{Numerical Results}
	\label{sec:numerical-results}
	
	In this section, we evaluate the performance of our proposed algorithm. The simulation parameters are as follows: 
	The number of antennas, $M$, is $100$. 
	The number of clients, $K$, is $20$. 
	The number of local iterations per communication round is $3$.
	The batch size is $500$.
	We perform image classification on the MNIST data set \cite{MNIST}.
	We have a very heterogeneous data distribution. Each client only has data from one digit: clients 1 and 2 have digit 0, clients 3 and 4 have digit 1, and so on. Each client has the same number of training samples.
	The network architecture is presented in Table~\ref{tab:net}. It has $d=12810$ trainable parameters.
	The clients keep $0.5\%$ of $\x_k^\text{full}$, i.e., $S = \lfloor 0.005 \, d / 2 \rfloor = 32$.
	The measurement matrices are constructed as, $\A = \frac{\A_r}{1.01\norm{\A_r}_2}$, where $\A_r$ has $\CN{0}{1}$ elements and $1.01$ is somewhat arbitrarily chosen to fulfill $\norm{\A}_2 < 1$. 
	The local learning rate is 0.01 and the global learning rate, $\alpha_t$, is $\frac{1}{3(10^{-4}t+1)}$.
	The large scale fading is equispaced (in dB scale) between $-40$~dB and $0$~dB, and the two clients with data of the same digit have equal $\beta_k$, i.e., $\beta_1 = \beta_2 = -40$~dB, $\dots$, $\beta_{19} = \beta_{20} = 0$~dB.
	The length of the transmitted vector is $T = 10S = 320$. 
	Finally, we vary the SNR, $\rho \in \{ 20, 30 \}$~dB and the number of pilot symbols, $\taup \in \{K, 10K\} = \{20,200\}$ (for both estimation methods). As mentioned in Section~\ref{sec:transmission}, we assume that the whole gradient update fits in a single coherence block. Since $T=320$, and $\taup=20$ or $\taup=200$ we need to have coherence intervals which are larger than $340$ or $520$ samples. In practice, $\tauc = 750$ in an outdoor vehicular scenario \cite[Ch.~2]{redbook}. 
	Since we assume a block fading model, the absolute bandwidth does not affect the system model.
	
	The sparsification is done as follows. If the clients are allowed to sparsify differently, they save only the $S$ elements of $\x_k^\text{full}$ with the highest magnitude. If the clients are forced to use the same pattern, one client is chosen, uniformly at random, and it decides the pattern for all clients~based~on~its~own~data.
	
	To solve problems \eqref{eq:opt} and \eqref{eq:optsum}, we implement iterative hard thresholding (IHT), which converges to a local optimum if $\norm{\A}_2 < 1$ \cite{Blumensath09}. We do a warm start in the solution of the matching pursuit algorithm \cite{Rish14}, where we stop once we have selected the desired number of non-zero components.

	\begin{figure*}
		\centering
		\includegraphics{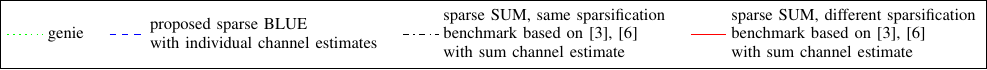}\\\vspace{-11pt}
		\subfloat{
			\includegraphics{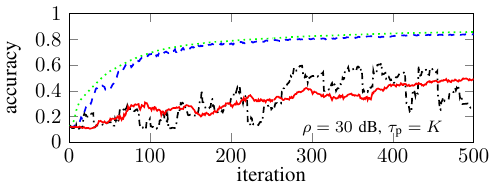}
		}
		\hfill 
		\subfloat{
			\includegraphics{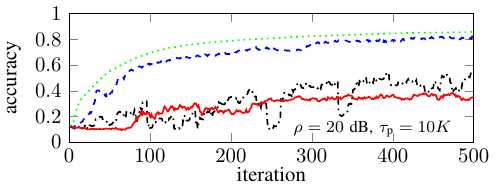}
		}
		\vspace{-10pt}
		\caption{Accuracy of MNIST digit classification with different receive combining methods.}
		\label{fig:results}
		\vspace{-5pt}
	\end{figure*}
	
	Fig.~\ref{fig:results} shows the test accuracy in two simulation scenarios where $\rho=30$~dB and $\taup=K$, and $\rho=20$~dB and $\taup=10K$, respectively. We simulate the proposed sparse BLUE algorithm (Algorithm~\ref{alg:sparse-blue}), where each client applies its own sparsity pattern, has a different measurement matrix, and uses full power. We also simulate two versions the benchmark sparse SUM algorithm (Algorithm~\ref{alg:sparse-sum}): one where the clients use the same sparsity pattern, and one where they use different sparsity patterns. In both these versions, the clients coordinate their power control according to \eqref{eq:power-constraint-sum}, and they use the same measurement matrix. For reference, we show the result of a genie (oracle) which has the full gradients from all clients.

	From the figures we can conclude that 
	\begin{enumerate}
		\item When the SNR is high, i.e., $\rho=30$~dB, the sparse BLUE performs very close to the genie even though only having $0.5\%$ of the gradients.
		\item Using different sparsification patterns in the sparse SUM algorithm gives a smaller variability in the accuracy.
		\item It is more accurate to estimate an $S$-sparse vector than a $KS$-sparse vector. The consequence is that in methods where the measurement matrix needs to be equal across clients, it is better for the clients to use the same sparsity pattern from the perspective of estimating the sparse vector. However, from a model convergence perspective, it is better for the clients to decide their own sparsity pattern, especially when the data is heterogeneous. This is a trade-off that needs to be made on a case-by-case basis. The advantage of the sparse BLUE method is that this problem is fully avoided.
		\item The overhead from channel estimation that is caused by individually estimating the clients' channels is compensated by the fact that the sparse BLUE method can tolerate smaller $T$ while still performing well. Here, $T$ is only $5\%$ of $d/2$, which significantly saves communication resources. Even though the channel estimation overhead can be reduced (by using a single pilot symbol) for the sparse SUM algorithm, it is not guaranteed that the $KS$ sparse vector can reliably be recovered from $T+(\taup-1)$ (more noisy) measurements.
	\end{enumerate}
	In simulations not shown here, using different measurement matrices does not significantly impact the performance.

	\section{Conclusions}
	
	We considered analog receive combining for federated edge learning with a MIMO receiver and no channel knowledge at the clients. We developed the optimal combiner (BLUE) if the parameter server has individual channel estimates to the clients. By using BLUE, we can separate the transmitted signals from each client which makes it possible for the clients to apply individual sparsification patterns of their gradients. 
	The sparsification allows for a significant reduction of the required communication resources and compensates for the communication overhead required for obtaining individual channel estimates. The proposed algorithm outperforms the benchmark OtA aggregation which uses a common pilot to estimate the sum channel \cite{Amiri21bfeel,Wei22}.
	Note that, the ``projection'' onto $\A$ followed by compressed sensing recovery is just \emph{one} way one can use to transmit the sparsified model updates.
	Finally, the sparse BLUE algorithm requires less coordination between clients than the benchmark and also allows for the use of robust aggregation rules and rejection algorithms. 
	
	\FloatBarrier

	\FloatBarrier

\end{document}